\documentclass[aps,prl,twocolumn,groupedaddress,showpacs]{revtex4-1}
\usepackage{graphicx}
\usepackage{amsmath}
\usepackage{textcomp}
\usepackage{mathrsfs}

\begin{document}

\title{Theory of Anisotropic Whispering Gallery Resonators}

\author{Marco Ornigotti}

\email{marco.ornigotti@mpl.mpg.de}

\author{Andrea Aiello}

\affiliation{Max Planck Institute for the Science of Light, G$\ddot{u}$nther-Scharowsky-Strasse 1/Bau24, 91058 Erlangen,
Germany} 

\date{\today}

\begin{abstract}
An analytic solution for an uniaxial spherical resonator is presented using the method of Debye potentials. This serves as a
starting point for the calculation of whispering gallery modes (WGM) in such a resonator. Suitable approximations for the radial
functions are discussed in order to best characterize WGMs. The characteristic equation and its asymptotic expansion for the
anisotropic case is also discussed, and an analytic formula with a precision of the order $O[\nu^{-1}]$ is also given. Our careful treatment of both boundary conditions and asymptotic expansions makes the present work a particularly suitable platform for a quantum theory of whispering gallery resonators.
\end{abstract}


\maketitle
\section{I. Introduction}
London's St. Paul's Cathedral is famous for its rich history and architecture; one of the most unique aspect of this building is
the whispering gallery that runs along the interior wall of its dome \cite{ref0}. When sounds are uttered in low voice against the
wall, sound waves generated circulate around the wall many times before fading away. As these waves propagate, they bring with
them sounds that are audible on the opposite side of the dome. On the contrary, if the same sounds are uttered at higher volume,
the frequencies of these sounds waves will not match and a lot of noise is created, making the message difficult to be heard at
any part of the wall.

The physical explanation of this effect was firstly given more than a century ago in terms of reflection of acoustic \emph{rays}
from a surface near the dome apex. It was initially assumed that the rays that propagate along different large arcs of the dome in
a form of a hemisphere should concentrate only at the point diametrically opposite to the source of the sound. Afterwards lord
Rayleigh, in his \emph{Theory of sound} \cite{ref1}, provided a different explanation of the effect that he named \emph{Whispering
Gallery Waves}: sound clutches to the wall surface and creeps along it without diverging as fast as during the free space
propagation: these sound waves then propagate within a narrow layer adjacent to the wall surface. It was then discovered, at the
beginning of the last century, that optical whispering gallery waves can exist even in dielectric spheres \cite{ref2,ref3}. An
optical resonator that shows this particular wave structure was then called \emph{Whispering Gallery Resonator} (WGR). In recent
times whispering gallery waves have found new fame with the development of nano-optics, in particular with the ability to
manufacture spherical and toroidal WGR with very high quality factors that ranges from $10^7$ to $10^{10}$
\cite{ref4bis,ref5bis,ref6bis}. This motivated a large theoretical and experimental work around these devices (see for example
Ref. \cite{ref4,ref5,ref6,ref7} and references therein). The ability to store light in microscopic spatial volumes for long
periods of time (due to the high $Q$-factor) resulted in a significant enhancement of nonlinear interactions of various kinds
like four-wave mixing \cite{ref8,ref9}, Raman \cite{ref10}, parametric and Brillouin scattering \cite{ref7,ref11}, microwave
up-conversion \cite{ref12}, second and third order harmonic generation \cite{ref13,ref14,ref15}. Besides the field of nonlinear
optics, WGRs were recently used even for cavity QED experiments \cite{ref15bis,ref15ter}. For an exhaustive review on the
applications of WGRs see Ref. \cite{ref15quater}.

Since many of the applications of these resonators involve nonlinear optics, WGR are commonly fabricated using nonlinear materials
or anisotropic crystals \cite{ref7}. Despite the wide scientific production in the theory of anisotropic spherical resonators, that
ranges from generalization of scattering methods \cite{ref17,ref18,ref18bis,ref18ter}, potential method \cite{ref17bis}, dyadic
Green function approach \cite{ref17ter} and Fourier-based analysis \cite{ref19}, and even though an extensive study of isotropic
WGRs was done in the past \cite{ref26}, detailed studies on anisotropic WGRs are still very few. To the knowledge of the authors
anisotropic WGRs are mainly reported in literature as studied with FDTD models \cite{ref20,ref21}, cavity loading \cite{ref22} and
direct solution of Maxwell's equation with a surface nonlinear polarization as a forcing term \cite{ref6}.

In this work, we intend to develop a suitable analytic theory for WGRs, starting from a review of the solutions of Maxwell's
equations in an uniaxial spherical resonator, then presenting and discussing its mode structure in the limit of small anisotropy,
and finally obtaining the spectrum of whispering gallery modes sustained by the resonator and their structure, discussing how
anisotropy influence those modes. A detailed discussion on the application of boundary conditions to this resonator is also
presented, pointing out how to apply correctly these conditions and discussing some of their basic features that,  to the
knowledge of the authors, is not present in earlier works. It is opinion of the authors that this discussion is important in
order to better understand the physics behind this problem.
This is the first main result of this work. 
 Finally, we introduce a more accurate approximation for the field
outside the resonator when the index of the Hankel function tends to infinity, as we noticed that the commonly used power expansion (as, for example,
the one presented in Ref. \cite{ref26}) does not match the exact function completely, i.e. it has an additional phase
factor respect to the real function. 
Such phase factor becomes relevant when field amplitude, as opposed to field intensity, turn to be fundamental. This happens, for example, when one wants to quantize the electromagnetic field inside the resonator, as required for a proper treatment of spontaneous emissions processes. Thus, the present work may serve as basis for a quantum theory of WGRs.
This is the second main result of this work.

This paper is organized as follows: in section II the Debye method of potentials for solving Maxwell's equations is briefly
presented and then used in section III to develop the theory of an anisotropic spherical resonator for a dielectric uniaxial
sphere. In section IV, \emph{Whispering Gallery Modes} (WGMs)  are obtained as limiting case of the normal modes of the dielectric
sphere with high quantum numbers and their spectrum is discussed.

\section{II. Debye method of potentials}
\subsection{A. Isotropic Solution}
Before considering the problem of an uniaxial spherical resonator, it is pedagogical to briefly review the method of Debye
potentials \cite{ref26}, largely used to solve Maxwell's equations in integrable systems.  Let us consider a monochromatic field
with an harmonic time dependence (i.e. , $\vec{E}(\vec{x},t)=\vec{E}(\vec{x})e^{-i\omega t}$) in an isotropic sourceless domain
$\Omega$; Maxwell's equations inside $\Omega$ can be written in the following symmetric form:\\
\begin{subequations}\label{maxeqGeneral}
\begin{eqnarray}
\nabla \times \vec{E} & = & -i k \vec{H},\\ \nabla \times \vec{H} & = & i k \vec{E} ,
\end{eqnarray}
\end{subequations}
where $k=\omega\sqrt{\varepsilon}/c$ is the wavevector in vacuum and and $\varepsilon$ is the dielectric constant inside $\Omega$.
In order to fully determine the fields, it is necessary to specify their values on the domain boundary $\partial\Omega$.
Electromagnetic boundaries are usually of two types: perfectly conducting walls (the field is zero on $\partial\Omega$), or open
systems, where the field components inside $\Omega$ and the ones outside $\Omega$ must match on $\partial\Omega$. The boundary,
together with symmetry considerations, gives a hint on which is the more suitable coordinate system to be used to solve the
problem (e.g. spherical coordinates for spheres, cilindrical coordinates for wires etc.) \cite{ref23}.

Let us specify our problem by considering an open system constituted by a sphere of dielectric constant $\varepsilon$ and radius
$R$ surrounded by an isotropic medium (i.e. air) \cite{ref24}. The set of Maxwell's equations \eqref{maxeqGeneral} in the
spherical reference frame can be written in the following compact form:\\
\begin{subequations}\label{Maxwell}
\begin{eqnarray}
\frac{\partial}{\partial\zeta_n}(L_mE_m)-\frac{\partial}{\partial\zeta_m}(L_nE_n) & = & - i k \epsilon_{lnm}L_n L_m
H_l,\label{maxwellA}\\ \frac{\partial}{\partial\zeta_n}(L_mH_m)-\frac{\partial}{\partial\zeta_m}(L_nH_n) & = & i k \epsilon_{lnm}
L_n L_m E_l,\label{maxwellB}
\end{eqnarray}
\end{subequations}
where $\{l,n,m\} \in  \{1, 2, 3\}$, $\zeta_m$ are the spherical coordinates ($\zeta_1=\varphi$, $\zeta_2=\theta$ and $\zeta_3=r$),
and  $L_m$ are the metric coefficients of the spherical reference frame ($L_1=r\sin\theta$, $L_2=r$, $L_3=1$). The Levi-Civita
symbol $\epsilon_{lnm}$ on the right-hand side of Eqs. \eqref{Maxwell} is equal to 1 if $\{l,n,m\}$ is equal to $\{1,2,3\}$ or any
of its even permutation, is equal to $-1$ for any odd permutation of $\{1,2,3\}$ and equal to zero elsewhere.

In this reference frame, the fields can be decomposed in the so-called  transverse electric (TE) and transverse magnetic (TM)
waves: TE waves are characterized by having $E_r=0$, i.e. the electric field is transveral with respect to the radial direction
$r$. TM waves are instead characterized by having the magnetic field transveral with respect to the radial direction (i.e.
$H_r=0$) \cite{ref23}. For the sake of simplicity, let us fix our attention on TM waves; the calculations for TE waves can be
straightforward obtained by analogy. From Eqs. \eqref{maxwellA} for $l=3$, by substituting $H_r=0$ it is possible to introduce the
$W$ function such that\\
\begin{subequations}\label{potW}
\begin{eqnarray}
r\sin\theta E_{\varphi} & = & \frac{\partial W}{\partial\varphi},\label{potWa}\\ rE_{\theta} & = & \frac{\partial
W}{\partial\theta}.\label{potWb}
\end{eqnarray}
\end{subequations}
By substituting relations \eqref{potW} into Eqs. \eqref{maxwellB} for $l=1,2$ and writing $W=\partial U/\partial r$, where $U$ is
the TM Debye potential, from \eqref{maxwellB} we obtain\\
\begin{subequations}
\begin{eqnarray}
H_{\varphi} & = & i k \frac{1}{r}\frac{\partial U}{\partial\theta},\\ H_{\theta} & = & - i k \frac{1}{r\sin\theta}\frac{\partial
U}{\partial\varphi},
\end{eqnarray}
\end{subequations}
and according to Eq \eqref{maxwellA}, the $r$-component ($l=3$) of the electric field is given by\\
\begin{equation}\label{EradialTM}
E_r = - \frac{1}{r^2\sin\theta}\Big[  \frac{\partial}{\partial\varphi}\Big( \frac{1}{\sin\theta}\frac{\partial}{\partial\varphi}
\Big) + \frac{\partial}{\partial\theta} \Big( \sin\theta\frac{\partial}{\partial\theta} \Big) \Big]U.
\end{equation}
Note that the differential operator that acts on the potential $U$ in this equation is the angular momentum operator
$\hat{\vec{L}}= - (\vec{r}\times\nabla)$, that is the same operator that originates the centrifugal potential in the Hydrogen atom
\cite{ref41}.

Therefore, all the components of the electric field are expressed in terms of the $U$ potential solely.
 In order to explicit them, it is necessary to find the equation which the $U$ potential satisfy. To do this, we can use one of
 the last two equations left available from Eq. \eqref{maxwellA}, i.e. the ones with $l=1,2$. By using one of them it is possible
 to obtain the following wave equation that $U$ must satisfy\\
\begin{equation}\label{waveU}
\frac{\partial^2 U}{\partial r^2}+ \nabla^2_{\perp}U + k^2U=0,
\end{equation}
where $\nabla^2_{\perp}$ is the angular part of the Laplace operator in spherical coordinates, i.e. the angular momentum operator
$\hat{\vec{L}}$.\\ The solution can be easily found with the method of separations of variables; writing the potential as
$U(r,\theta,\varphi)=\Psi(r)\Theta(\theta)\Phi(\varphi)$ and substituting this into Eq. \eqref{waveU}, we obtain the following
equations for the functions $\Psi(r)$, $\Theta(\theta)$ and $\Phi(\varphi)$ \cite{ref23}:\\
\begin{subequations}
\begin{eqnarray}
\frac{d^2\Psi}{dr^2} & + & \Big( k^2-\frac{c_3}{r^2} \Big)\Psi = 0,\label{parteRadiale}\\ \frac{1}{\sin\theta}\frac{d}{d\theta}
\Big( \sin\theta\frac{d\Theta}{d\theta} \Big) & + &  \Big( c_1 - \frac{c_2}{\sin^2\theta} \Big)\Theta = 0,\label{parteTheta}\\
\frac{d^2\Phi}{d\varphi^2} & + & c_2\Phi = 0,\label{partePhi}
\end{eqnarray}
\end{subequations}
where $c_1$, $c_2$ and $c_3$ are the separation constants appearing in the equations by separating the variables, whose values
must be $c_1=n(n+1)$ and $c_2=m^2$ in order to the solution to these equations to be unique, i.e. physically meaningful; $n$ and
$m$ are integers, including zero. With these values the solution is straightforward. Equations \eqref{parteTheta} and
\eqref{partePhi} give rise to the so-called spherical harmonics
$Y_{nm}(\theta,\varphi)=\mathscr{N}P_n^m(\cos\theta)e^{im\varphi}$, i.e. the eigensolutions of the angular momentum operator
\cite{ref23}, where $P_n^m(\cos\theta)$ are the associated Legendre functions of the first kind that are solution of Eq.
\eqref{parteTheta}, while the complex exponential is a solution of Eq. \eqref{partePhi}. $\mathscr{N}$ is a normalization constant
that guarantees that the integral over the solid angle is unitary. The radial equation \eqref{parteRadiale} can be transformed
into a Bessel equation by the substitution $\Psi(r)=\sqrt{kr}Z(kr)$ that brings to:\\
\begin{equation}\label{besselRadiale}
\frac{d^2Z}{dx^2}+\frac{1}{x}\frac{dZ}{dx}+ \Big( 1- \frac{\nu^2}{x^2} \Big)Z=0,
\end{equation}
where $x=kr$ and $\nu=n+1/2$; the solutions to this equation are the four Bessel functions $J_{\nu}(x)$, $N_{\nu}(x)$,
$H^{(1)}_{\nu}(x)=J_{\nu}(x)+iN_{\nu}(x)$ and $H^{(2)}_{\nu}(x)=J_{\nu}(x)-iN_{\nu}(x)$. Physically, the solution inside the
sphere must be finite at the origin, and the only plausible solution is $J_{\nu}(x)$ because $N_{\nu}(x)$ has a divergence at the
origin. Outside the sphere, instead, the solution should have the form of a runaway wave with the Sommerfeld condition at the
infinity (i.e. , the solution must drop at infinity as the inverse square of the distance). For this reason the correct solution
in this domain is the Hankel function of the first kind $H^{(1)}_{\nu}(x)$ because its asymptotic form decreases to zero as the
inverse square of the distance for $x\rightarrow\infty$. Putting everything together, the TM Debye potential reads as follows:\\
\begin{equation}\label{solution}
U_{nm}^{int/ext}(r,\theta,\varphi) = C_{int/ext} \sqrt{k r}Z_{\nu}(k r)Y_{nm}(\theta,\varphi),
\end{equation}
where $n,m$ are the angular quantum numbers that address the single mode of the resonator, $Z_{\nu}(k r)$ is the radial
Bessel-type function that is equal to the Bessel function $J_{\nu}(k r)$ inside the dielectric sphere, and is equal to the Hankel
function of the first kind $H_{\nu}^{(1)}(k_0 r)$ outside the dielectric sphere. The constants $C_{int/ext}$ are to be determined
by applying suitable boundary conditions. Note that the argument of the Bessel function inside the sphere contains the sphere
dielectric constant $\varepsilon$ via the wavevector $k=\omega\sqrt{\varepsilon}/c$ while the argument of the Hankel function
outside the sphere contains only the vacuum wavevector $k_0=\omega/c$ because $\varepsilon_{air}=1$.\\ The components of the
electric and magnetic fields for TM waves can then be written as a function of $U$ as follows:\\
\begin{subequations}\label{TMcomponent}
\begin{eqnarray}
E_r & = & \Big( \frac{\partial^2}{\partial r^2}+k^2 \Big)U,\\ E_{\theta} & = & \frac{1}{r}\frac{\partial^2 U}{\partial
r\partial\theta},\\ E_{\varphi} & = & \frac{1}{r\sin\theta}\frac{\partial^2 U}{\partial r \partial\varphi},\\ H_r & = & 0,\\
H_{\theta} & = & -ik\frac{1}{r}\frac{\partial U}{\partial\varphi},\\ H_{\varphi} & = & ik\frac{1}{r}\frac{\partial
U}{\partial\theta}.
\end{eqnarray}
\end{subequations}
Note that in obtaining the expression of $E_r$ we have combined Eqs.\eqref{EradialTM} and \eqref{waveU}.\\ If we proceed in a
similar manner for TE waves, we obtain:\\
\begin{subequations}\label{TEcomponent}
\begin{eqnarray}
E_r & = & 0,\\ E_{\theta} & = & -ik\frac{1}{r}\frac{\partial V}{\partial\varphi},\\ E_{\varphi} & = & ik\frac{1}{r}\frac{\partial
V}{\partial\theta},\\ H_r & = & \Big( \frac{\partial^2}{\partial r^2}+k^2 \Big)V,\\ H_{\theta} & = & \frac{1}{r}\frac{\partial^2
V}{\partial r\partial\theta},\\ H_{\varphi} & = & \frac{1}{r\sin\theta}\frac{\partial^2 V}{\partial r \partial\varphi},
\end{eqnarray}
\end{subequations}
where $V$ is the TE field potential obtained by Eqs. \eqref{Maxwell} by substituting the TE ansatz $E_r=0$. 
\subsection{B. Boundary Conditions}
Prior to investigate the structure of the modes for the anisotropic resonator, it is important to discuss the boundary conditions
that have to be applied to this problem. At the resonator surface $r=R$, the wave vector $k$ inside the dielectric sphere has to
match the wave vector $k_0=\omega/c$ outside the sphere and the constants $C_{int}$ and $C_{ext}$ should be chosen properly.\\
There is not a unique way to fulfill boundary conditions: in fact, one could apply ``pure" or ``mixed" conditions: the former
consists in applying the boundary conditions to all the components of only electric \emph{or} magnetic field, while the latter
applies the boundary to certain components of one field and certain other components of the other field. Obviously, these two
different paths bring to the same physical solutions \cite{ref23}.  Among these possibilities, in this work we chose to apply
``pure" boundary condition, i.e. we impose that the tangential electric (magnetic) field components for TM (TE) waves has to be
continuous at the resonator surface $r=R$, while the radial component of the displacement vector $\vec{D}=\varepsilon\vec{E}$ is
continuous across the resonator surface. For TE fields, the radial condition is automatically fulfilled, since the resonator is
non-magnetic (i.e. , $\mu=1$).

The condition for the radial component of the displacement vector ($\varepsilon_{int}E_r^{int}=\varepsilon_{ext}E_r^{ext}$) across
the resonator surface gives the ratio between the inner and outer coefficients\\
\begin{equation}
\frac{C_{ext}}{C_{int}}=\varepsilon^{1/4}\frac{j_{\nu}(k_0\sqrt{\varepsilon}R)}{h_{\nu}^{(1)}(k_0 R)},
\end{equation}
while the continuity of the tangential component $E_{\theta,\varphi}^{int}=E_{\theta,\varphi}^{ext}$ of the field gives rise to
the so called characteristic equation, that allows to determine the allowed values for the wave vector $k$ (i.e. , to find the
spectrum of the allowed modes) inside the resonator, and it turns out to be\\
\begin{equation}\label{eigenTM}
\frac{[j_{\nu}(kR)]'}{j_{\nu}(kR)}=\sqrt{\varepsilon} \frac{[h_{\nu}^{(1)}(k_0R)]'}{h_{\nu}^{(1)}(k_0R)},
\end{equation}
for TM waves and
\begin{equation}\label{eigenTE}
\frac{[j_{\nu}(kR)]'}{j_{\nu}(kR)}=\frac{1}{\sqrt{\varepsilon}} \frac{[h_{\nu}^{(1)}(k_0R)]'}{h_{\nu}^{(1)}(k_0R)},
\end{equation}
for TE waves. In these equations $j_{\nu}(x)=\sqrt{x}J_{\nu}(x)$ and $h_{\nu}^{(1)}(x)=\sqrt{x}H_{\nu}^{(1)}(x)$ are the
Riccati-Bessel functions, and the prime indicates the total derivative with respect to the argument on which the functions depend,
i.e. over $kR$ or $k_0R$.

Although formally corrected, as they are these boundary conditions do not provide a unique solution to the determination of the
mode patterns in the resonator.

In order to better understand this not-uniqueness of the solution, let us consider the general structure of eqs. \eqref{eigenTM}
and \eqref{eigenTE}. Let $T(x)$ be a piecewise function defined across an interface, placed at $x=1$, between two region of space,
such that $T(x)=C^{(i)}f(x)$ for $x<1$ and $T(x)=C^{(e)}g(x)$ for $x>1$, with $f(x)$ and $g(x)$ two arbitrary real valued and
regular functions. The constants $C^{(i,e)}$ are to be determined by the boundary conditions and they must be chosen in such a way
that the following characteristic equation is satisfied:\\
\begin{equation}\label{generalB}
\frac{f'(x)}{f(x)}=\alpha\frac{g'(x)}{g(x)},
\end{equation}
where the apex stands for the derivative of the two functions with respect to their arguments. Since this is a generalization of
the characteristic equations \eqref{eigenTM} and \eqref{eigenTE}, this equation must hold at the interface between the two region
of space considered, i.e. its validity is limited to $x=1$. From eq. \eqref{generalB} it is clear that if we admit that the
derivatives $f'(x)$ and $g'(x)$ of the functions are equal at the separation interface $x=1$, then the functions themselves will
be discontinuous with a jump that has the value of $1/\alpha$ . On the other hand if we now admit that the the functions $f(x)$
and $g(x)$ are equal at the separation interface $x=1$, then their derivatives must be discontinuous, and the magnitude of the
discontinuity is precisely $\alpha$.

\begin{figure}[!t] 
\includegraphics[width=0.5\textwidth]{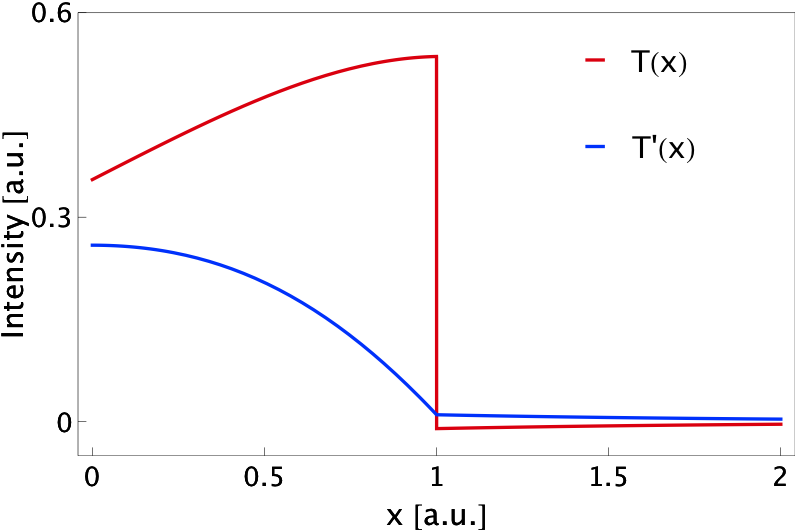} 
\caption{(color online) The figure shows the behavior
of the function $T(x)$ and its derivative $T'(x)$ when the condition of continuous derivative at the separation interface $x=1$ is
considered. As can be noted, in this case the function shows a discontinuity while its derivative is (obviously) continuous. For
this example we have used $f(x)=\mathrm{Ai}(x)$ and $g(x)=e^{-x}$, and the magnitude of the discontinuity in the function $T(x)$
at the separation interface is $f'(1)/g'(1) \simeq 0.5457$.} 
\label{figuraBoundary1} 
\end{figure}

The first situation corresponds to require that the derivative of the function $T(x)$ is continuous at the separation interface
(i.e. $T'(1)^+ = T'(1)^-$, where the plus or minus superscript stands for the expression of $T(x)$ for $x>1$ and $x<1$
respectively). This implies that $C^{(e)}/C^{(i)} = f'(1)/g'(1)$ and the function $T(x)$ can be written as:\\
 \begin{displaymath}
T(x) = \left\{
\begin{array}{lc}
\displaystyle{f(x)} & \displaystyle{x<1}, \\
\\
\\
\displaystyle{\Bigg[ \frac{f'(1)}{g'(1)} \Bigg] g(x)} & \displaystyle{x\geq 1}.
\end{array}
\right.
\end{displaymath}
\vspace{2mm}\\ It is then clear that taking $T'(x)$ to be continuous at the interface results in a discontinuity in the behavior
of $T(x)$ while passing through $x=1$, whose magnitude is $f'(1)/g'(1)$, as is depicted in Fig.\ref{figuraBoundary1}.

\begin{figure}[!t]
\includegraphics[width=0.5\textwidth]{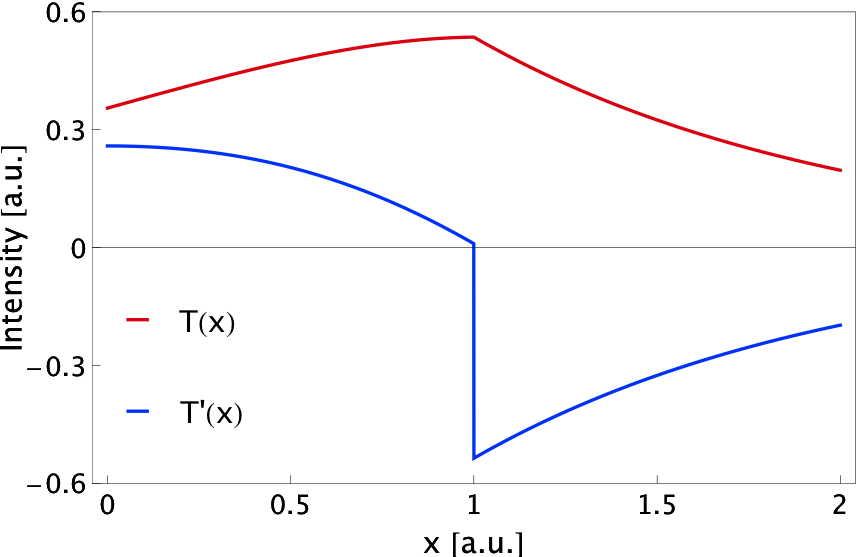} 
\caption{(color online) The figure shows the behavior of the function
$T(x)$ and its derivative $T'(x)$ when the condition of continuous function at the separation interface $x=1$ is considered. As
can be noted, in this case the function itself is (obviously) continuous while its derivative shows a jump discontinuity. For this
example we have used $f(x)=\mathrm{Ai}(x)$ and $g(x)=e^{-x}$ and the magnitude of the discontinuity in the derivative $T'(x)$ at
the separation interface is $f(1)/g(1) \simeq 0.5457$.} 
\label{figuraBoundary2} 
\end{figure}

Conversely, the second condition on the functions $f(x)$ and $g(x)$ implies that the function $T(x)$ to be continuous at the
separation interface (i.e. , $T(1)^+ = T(1)^-$), we have $C^{(e)}/C^{(i)} = f(1)/g(1)$ and the function $T(x)$ has the following
form:\\
\begin{displaymath}
T(x) = \left\{
\begin{array}{lc}
\displaystyle{f(x)} & \displaystyle{x<1}, \\
\\
\\
\displaystyle{\Bigg[ \frac{f(1)}{g(1)} \Bigg]g(x)} & \displaystyle{x\geq 1}.
\end{array}
\right.
\end{displaymath}
\vspace{2mm}\\ In this case, taking $T(x)$ to be continuous at the interface results in a discontinuity in its derivative, whose
magnitude is $f(1)/g(1)$, as Fig.\ref{figuraBoundary2} underlines. 

In both cases, however, it is not possible to make the functions \emph{and} the derivatives both continuous at the same time. This
fact makes only possible to obtain the ratio between the two constants $C_{int}, C_{ext}$ and not their explicit value: in order
to do that, another condition must be applied to the problem. This condition depends on the particular problem we are dealing on;
in scattering problems, for example, the incoming field is known, and determines the field pattern on the resonator surface. In
this case $C_{ext}$ is known and the ambiguity is removed. Another situation in which the ambiguity is overcome is by embedding
the whole system (resonator plus surrounding medium) in an ideal perfectly reflective sphere of big, but finite radius $R_0$, in
such a way that the boundary conditions at the metallic surface will completely determine the fields: this second approach is very
useful if we are dealing with the quantization of the field in such a system.

We want to end this discussion by pointing out that the first situation (derivative continuous at the interface) corresponds to
the boundary condition for the electric field across a dielectric surface: the normal component with respect to the separation
surface is discontinuous by a factor equal to the ratio of the two dielectric constants of the two regions, while the tangential
components (i.e. , the derivative of the radial field in our spherical case) is continuous at the interface. The second situation,
instead, corresponds to put continuous the normal component of the displacement vector across the separation surface, resulting in
a discontinuity of the tangential component of the displacement vector at the interface. While the former correspond to the usual
way of imposing boundary conditions in an electromagnetic problem, the latter is never used, but still valid.

In this work, however, we are neither interested on scattering problems nor on field quantization, and so in the rest of the paper
this ambiguity will not be removed. This does not create too much problems because we are only interested on the mode structure of
the resonator. We leave this problem of not-uniqueness to future works.\\
\section{III. Normal modes of an uniaxial spherical resonator}
Let us consider the same dielectric spherical resonator of radius $R$ of the previous section, but with an uniaxial anisotropy
along the $z$-axis described by the following dielectric tensor:\\
\begin{eqnarray}\label{epsilon}
\hat{\varepsilon}& =& \left(
\begin{array}{ccc}
\varepsilon_{xx} & 0 & 0\\ 0 & \varepsilon_{xx} & 0\\ 0 & 0 & \varepsilon_{zz}\\
\end{array}
\right)= \nonumber \\ \vspace{1mm}
 & = &\varepsilon_{xx} (\mathbf{ \hat{x}\hat{x}}+\mathbf{\hat{y}\hat{y}})+\varepsilon_{zz}\mathbf{\hat{z}\hat{z}}.
\end{eqnarray}
In order to use this dielectric tensor in Eqs. \eqref{Maxwell}, it should be converted in spherical coordinates; this operation is
simply done by converting the cartesian dyadics $\mathbf{\hat{x}\hat{x}}, \mathbf{\hat{y}\hat{y}}$ and $ \mathbf{\hat{z}\hat{z}}$
into the spherical dyadics $\mathbf{\hat{r}\hat{r}}, \boldsymbol{\hat{\theta}\hat{\theta}}$ and
$\boldsymbol{\hat{\varphi}\hat{\varphi}}$ using the standard cartesian-to-spherical transformation relations \cite{ref23}. By
performing this transformation, the dielectric tensor in spherical coordinates reads\\
\begin{equation}
\hat{\varepsilon}=\left(
\begin{array}{ccc}
\varepsilon_{rr} & -\varepsilon_{r\theta} & 0\\ -\varepsilon_{r\theta} & \varepsilon_{\theta\theta} & 0\\ 0 & 0 &
\varepsilon_{\perp}\\
\end{array}
\right),
\end{equation}
with $\varepsilon_{\pm} =(\varepsilon_{zz}\pm\varepsilon_{xx})/2$ and $\varepsilon_{\perp}=\varepsilon_{xx}$. We have then defined
$\varepsilon_{rr}=\varepsilon_+ +\varepsilon_- \cos(2\theta)$, $\varepsilon_{\theta\theta}= \varepsilon_+ - \varepsilon_-
\cos(2\theta)$ and $\varepsilon_{r\theta}=\varepsilon_- \sin(2\theta)$. The fact that the tensor components depend on the polar
coordinate $\theta$ makes the problem to find the eigenmodes of the spherical resonator much more difficult. Moreover, in an
anisotropic system it is in general no longer possible to divide the electric and magnetic fields in their TM and TE components.
In order to overcome the latter problem, we will focus our attention on the case of small anisotropy, i.e.
$\lambda=\varepsilon_-/\varepsilon_+\ll 1$ (this approximation is very good if we consider, for example, a dielectric sphere made
of Lithium Niobate ($\mathrm{LiNbO_3}$) for which we have $\varepsilon_{xx}=5.3$, $\varepsilon_{zz}= 6.47$ and therefore
$\lambda\simeq 0.01$). In such a way the fields can be decomposed in quasi-TE and quasi-TM oscillations, allowing us to solve the
problem using the method of Debye potentials illustrated above. To this aim, and for the sake of clearness, let us rewrite the set
of Eqs. \eqref{Maxwell} for the anisotropic case as follows:\\
\begin{subequations}\label{AniMaxwell}
\begin{eqnarray}
\frac{\partial}{\partial r}(r\sin\theta E_{\varphi})-\frac{\partial}{\partial\varphi}(E_r) & = & ik_0 \sin\theta
(rH_{\theta}),\label{prima}\\ \frac{\partial}{\partial\theta}(E_r)-\frac{\partial}{\partial r}(rE_{\theta}) & = & ik_0
(rH_{\varphi}),\label{seconda}\\ \frac{\partial}{\partial\varphi}(r E_{\theta})-\frac{\partial}{\partial\theta}(r\sin\theta
E_{\varphi}) & = & ik_0 r\sin\theta (rH_r),\label{terza}
\end{eqnarray}
\end{subequations}
and\\
\begin{subequations}\label{AniMaxwell2}
\begin{eqnarray}\label{quarta}
\frac{\partial}{\partial r}&(&r\sin\theta H_{\varphi})-\frac{\partial}{\partial\varphi}(H_r) = \nonumber \\
 & = &  ik_0 \sin\theta[ \varepsilon_{r\theta}(rE_r) - \varepsilon_{\theta\theta} (rE_{\theta})],
\end{eqnarray}
\begin{equation}\label{quinta}
\frac{\partial}{\partial\theta}(H_r)-\frac{\partial}{\partial r}(rH_{\theta}) =  -ik_0 \varepsilon_{\perp} (rE_{\varphi}),
\end{equation}
\begin{eqnarray}\label{sesta}
\frac{\partial}{\partial\varphi}(r H_{\theta})& -& \frac{\partial}{\partial\theta}(\sin\theta rH_{\varphi})=  \nonumber \\ & = &-
ik_0 r\sin\theta [\varepsilon_{rr}( rE_r) +\varepsilon_{r\theta} (rE_{\theta})].
\end{eqnarray}
\end{subequations}
As in the previous section, we solve the problem for the quasi-TM component of the field (i.e. $H_r=0$); the quasi-TE solution is
again obtained using similar arguments. We follow the solving procedure described in Ref.\cite{ref29}. Equation \eqref{terza}
defines the $W$ function as in \eqref{potW}. Let us combine \eqref{quinta} and the derivative with respect to $r$ of
\eqref{prima}:\\
\begin{displaymath}
\left\{
\begin{array}{l}
\displaystyle{\frac{\partial}{\partial r}(rH_{\theta})  = \frac{1}{\sin\theta}ik_0\varepsilon_{\perp}\frac{\partial
W}{\partial\varphi},} \\
\\
\\
\displaystyle{\frac{\partial^3}{\partial r^2\partial\varphi} (W) - \frac{\partial^2}{\partial r\partial\varphi}(E_r) =
ik_0\sin\theta\frac{\partial}{\partial r}(rH_{\theta}).}
\end{array}
\right.
\end{displaymath}
\vspace{2mm}\\ If we substitute the expression of $\frac{\partial}{\partial r}(rH_{\theta})$ obtained from the first equation into
the second one and if we define the differential operator $\hat{l}_x=\partial^2/\partial r^2 + k_0^2\varepsilon_{\perp}$ we
obtain:\\
\begin{equation}\label{radial}
\frac{\partial E_r}{\partial r}=\hat{l}_x W.
\end{equation}
Combining now \eqref{quarta}, and the derivative with respect to $r$ of \eqref{seconda} gives:\\
\begin{displaymath}
\left\{
\begin{array}{l}
\displaystyle{\frac{\partial}{\partial r}(r\sin\theta H_{\varphi})  = ik_0r\sin\theta \Big( \varepsilon_{r\theta}E_r -
\frac{1}{r}\varepsilon_{\theta\theta}\frac{\partial W}{\partial\theta} \Big),} \\
\\
\\
\displaystyle{\frac{\partial^2}{\partial r\partial\theta}(E_r) - \frac{\partial^2}{\partial r^2} (rE_{\theta}) =
ik_0\frac{\partial}{\partial r} \Big( rH_{\varphi} \Big).}\\
\end{array}
\right.
\end{displaymath}
Again, by substituting the expression for $\frac{\partial}{\partial r}(rH_{\varphi})$ obtained from the first equation into the
second one, and by defining the differential operator $\hat{l}_{\theta}=\partial^2/\partial r^2 + k_0^2\varepsilon_{\theta\theta}$
we obtain\\
 \begin{equation}\label{intermedia}
\Big[ k_0^2\varepsilon_{r\theta}r+\frac{\partial^2}{\partial r\partial\theta} \Big]E_r=\hat{l}_{\theta}\frac{\partial
W}{\partial\theta}.
\end{equation}
As in the isotropic case, we want to define $W$ as a function of the quasi-TM potential $U$, in order to fully determine the
components of the fields as a function of the quasi-TM potential $U$ solely. In order to do this, let us compare Eqs.
\eqref{radial} and \eqref{intermedia}. By noting that the differential operator $\hat{l}_{\theta}$ commutes with the operator
$(k_0^2 \varepsilon_{r\theta}r + \partial^2/\partial r\partial\theta)$ that appears in \eqref{intermedia}, is it possible to
define, after some simple algebra, the function $W$ as a function of the quasi-TM potential $U$ as follows:\\
\begin{equation}
W=\frac{\partial}{\partial r} \Big( \hat{l}_{\theta} U \Big).
\end{equation}
This allow us to write the components of the TM electric and magnetic field in terms of the quasi-TM potential $U$ as follows
\cite{ref28}:\\
\begin{subequations}\label{TMfield}
\begin{eqnarray}
E_r ^{TM}  & = & \hat{ l}_x\hat{ l}_{\theta} U,\\ rE_{\varphi}^{TM} & = & \frac{1}{\sin\theta}\frac{\partial^2}{\partial
r\partial\varphi}(\hat{l}_{\theta}U),\\ rE_{\theta}^{TM} & = & \Big( \varepsilon_{r\theta}k_o^2 r + \frac{\partial^2}{\partial
r\partial\theta} \Big) \hat{l}_x U,\\ H_r^{TM} & = & 0,\\ rH_{\theta}^{TM} & = & \frac{i k_0
\varepsilon_{\perp}}{\sin\theta}\frac{\partial}{\partial\varphi}(\hat{l}_{\theta}U),\\ rH_{\varphi}^{TM} & = & i k_0 \Big[
\varepsilon_{\theta\theta}\frac{\partial}{\partial\theta} + \varepsilon_{r\theta} \Big( 1-r\frac{\partial}{\partial r} \Big) \Big]
\hat{l}_x U.
\end{eqnarray}
\end{subequations}
Again, if we proceed in a similar manner for the quasi-TE waves we obtain:\\
\begin{subequations}\label{TEfield}
\begin{eqnarray}
H_r^{TE} & = & \hat{l}_x\hat{l}_{\theta} V,\\ rH_{\theta}^{TE} & = & \frac{\partial^2}{\partial
r\partial\theta}(\hat{l}_{\theta}V),\\ rH_{\varphi}^{TE}& = & \frac{1}{\sin\theta}\frac{\partial^2}{\partial
r\partial\varphi}(\hat{l}_x V),\\ E_r ^{TE} & = & 0,\\ rE_{\theta}^{TE} & = &
\frac{ik_0}{\sin\theta}\frac{\partial}{\partial\varphi}(\hat{l}_x V),\\ rE_{\varphi}^{TE} & = &
-ik_0\frac{\partial}{\partial\theta}(\hat{l}_x V),
\end{eqnarray}
\end{subequations}
where $V$ represents the quasi-TE potential.\\ Equations \eqref{TMfield} and \eqref{TEfield} represent the quasi-TM and quasi-TE
components of electric and magnetic field inside an anisotropic spherical resonator in terms of the TM and TE quasi-potentials $U$
and $V$.\\ The next step consists in constructing an equation that the quasi-potentials $U$ and $V$ satisfy, whose solutions give
the mode fields of the resonator. To do that, let us consider a general electric and magnetic field, whose components are written
as the superposition of the quasi-TE and quasi-TM oscillations, i.e. $E_i=E_i^{TM}+E_i^{TE}$ and $H_i=H_i^{TM}+H_i^{TE}$.
Substituting this ansatz in Eqs.  and \eqref{AniMaxwell2}, after some algebra we arrive at a set of two coupled equations for the
quasi-potentials $U$ and $V$ that reads \cite{ref29}\\
\begin{subequations}\label{coupled}
\begin{equation}\label{coupledU}
\hat{L}_H V = 2 i \varepsilon_- k_0 \hat{\Upsilon}\frac{\partial U}{\partial\varphi},
\end{equation}
\begin{equation}\label{coupledV}
\hat{L}_E U = 2 i \varepsilon_- k_0\hat{\Upsilon}\frac{\partial V}{\partial\varphi},
\end{equation}
\end{subequations}
where we have defined:\\
\begin{subequations}
\begin{eqnarray}
\hat{L}_H & = & (\nabla^2_{\perp}+r^2 \hat{l}_x)\hat{l}_{\theta}-2\varepsilon_-
k_0^2\frac{\partial^2}{\partial\varphi^2},\nonumber\\ \hat{L}_E & = & \hat{\Xi}_0 + \lambda\hat{\Xi}\hat{l}_x,\nonumber\\
\nabla^2_{\perp} & = &
\frac{1}{\sin\theta}\frac{\partial}{\partial\theta}\sin\theta\frac{\partial}{\partial\theta}+\frac{1}{\sin^2\theta}\frac{\partial^2}{\partial\varphi^2},\nonumber\\
\hat{\Upsilon} & = & \Big(  r \hat{l}_x -2\frac{\partial}{\partial r} \Big)\cos\theta
-\sin\theta\frac{\partial^2}{\partial\theta\partial r},\nonumber\\ \hat{\Xi}_0 & = & \Big[ \nabla^2_{\perp} +
r^2\frac{\partial^2}{\partial r^2}+ \varepsilon_+ \Big(1-\lambda^2 \Big)k_0^2r^2 \Big]\hat{l}_x +\nonumber \\ & + & 2\varepsilon_-
(1-\lambda)k_0^2\frac{\partial^2}{\partial\varphi^2},\nonumber\\ \hat{\Xi} & = & \cos(2\theta) \Big[ r^2\frac{\partial^2}{\partial
r^2} - \nabla^2_{\perp} + 3 \Big(1-r\frac{\partial}{\partial r} \Big) \Big] +\nonumber\\
 & + & \Big( 3-2r\frac{\partial}{\partial r} \Big) \sin(2\theta)\frac{\partial}{\partial\theta} +\nonumber \\
 & + & \Big( 1-r\frac{\partial}{\partial r} \Big)-2\frac{\partial^2}{\partial\varphi^2},\nonumber
\end{eqnarray}
\end{subequations}
and $\lambda=\varepsilon_-/\varepsilon_+$ is the anisotropy parameter.

From Eqs. \eqref{coupled} it is evident that the anisotropy gives rise to a coupling between the two quasi-potentials $U$ and $V$;
this coupling is  absent in the isotropic case in which the two potentials are independent one each other. These equations, in
fact, contain the isotropic solution in the limit of $\lambda=0$. Making this substitution in equation \eqref{coupledU} and using
the definition of the operator $\hat{L}_H$, we obtain:\\
\begin{equation}
(\nabla^2_{\perp}+r^2\hat{l}_x)\hat{l}_{\theta}U=0.
\end{equation}
Because we set $\lambda=0$, the two differential operators $\hat{l}_x$ and $\hat{l}_{\theta}$ are equal, since
$\varepsilon_{\theta\theta}=\varepsilon_+ - \varepsilon_-\cos2\theta = \varepsilon_+ =\varepsilon$ and
$\varepsilon_{\perp}=\varepsilon$, i.e. no anisotropy is present anymore. We now define $\hat{l}_{\theta}U=A$ as the isotropic
potential, and we assume that this potential can be written in a separable way, i.e.
$A(r,\theta,\varphi)=\Psi(r)Y_{nm}(\theta,\varphi)$, where $Y_{nm}(\theta,\varphi)$ are the eigensolutions of the angular momentum
operator, whose eigenvalues are $-n(n+1)$ (i.e. , $\nabla^2_{\perp}Y_{nm}=-n(n+1)Y_{nm}$). Substitution of this ansatz into the
previous equation and consequent simplification of the angular part then brings to the following radial equation:\\
\begin{equation}\label{reduceTo}
[n(n+1)-r^2\hat{l}_x]\Psi(r)=0,
\end{equation}
that is precisely the radial equation \eqref{parteRadiale} for the isotropic potential, whose solutions are the Bessel functions
given in the previous section. The same procedure applied to the quasi-potential $V$ brings to its isotropic counterpart.\\ Since
we stated that the anisotropy is small (i.e. , $\lambda\ll 1$), then we can use the method of separations of variable to solve the
coupled equations \eqref{coupled}. We then write the quasi-potentials as follows:\\
\begin{subequations}\label{ansatz}
\begin{equation}
U(r,\theta,\varphi)=\sum_p U_p(r,\theta,\varphi)=\sum_p u_p(r)Y_{nm}(\theta,\varphi),
\end{equation}
\begin{equation}
V(r,\theta,\varphi)=\sum_p V_p(r,\theta,\varphi)=\sum_p v_p(r)Y_{nm}(\theta,\varphi),
\end{equation}
\end{subequations}
where the subscript $p$ stands for the three indexes $n$,$m$ and $q$ on which the quasi-potential depends; the polar index $n$
determines the number of field nodes along the polar coordinate $\theta$, the azimuthal number $m$ characterizes the nodes in the
$\varphi$ direction and, finally, the radial index $q$ gives the number of field oscillations along the radial direction $r$ that
is related with the solution of the characteristic equation. Substituting \eqref{ansatz} into \eqref{coupled}, using relationships
(4) and (5) of Ref.\cite{ref29} and equating the terms with equal angular part $Y_{nm}(\theta,\varphi)$, we obtain the following
set of differential equations for the radial components $u_p(r)$ and $v_p(r)$ of the quasi-potentials\cite{ref30}:\\
\begin{subequations}\label{radialpart}
\begin{eqnarray}
a_{1,n}^{TM} u_n & - & a_{2,n-2}^{TM} u_{n-2} - a_{3,n+2}^{TM} u_{n+2} =\nonumber \\  & - & 2\varepsilon_- mk_0[ b_{1,n-1} v_{n-1}
- b_{2,n+1} v_{n+1} ],
\end{eqnarray}
\begin{eqnarray}
a_{1,n}^{TE} v_n & - & \lambda(a_{2,n-2}^{TE} v_{n-2} + a_{3,n+2}^{TE} v_{n+2}) =\nonumber \\ & - & 2\lambda mk_0[b_{1,n-1}
u_{n-1}- b_{2,n+1} u_{n+1}].
\end{eqnarray}
\end{subequations}
For the sake of clarity, the expresison of the coefficients $a_i^{TM/TE}$ and $b_i$ are reported in Appendix A.

These equations require, in general, a numerical approach to be solved. However, in the limit of small anisotropy, i.e. $\lambda
\ll 1$, a solution to Eqs. \eqref{radialpart} can be searched in terms of power series in the factor $\lambda$. The zeroth order
solution brings (as shown before) to the solution of the isotropic spherical resonator in terms of the Riccati-Bessel functions
[see Eq. \eqref{reduceTo}]. The first order solution, i.e. the anisotropic correction we are searching for, is obtained by
neglecting the terms that are proportional to $\lambda^2$ in Eqs. \eqref{radialpart}: however, the resulting equations contain in
the right-hand side a term that is not zero (like in the zeroth order solution) but depends on the quasi-potentials $u_{n\pm1}$
and $v_{n\pm1}$. This coupling among neighbor radial modes is a signature of the anisotropy, that on one hand breaks the azimuthal
degeneracy [the azimuthal quantum number appears in the definition of the coefficients of Eqs. \eqref{radialpart}] and on the
other hand results in a coupling between radial modes. Although this coupling results in an impossibility of an analytic solution,
it can be demonstrated \cite{ref29} that these terms are of the order $\lambda^2$ and at the first order they can be neglected.
With this argument, Eqs. \eqref{radialpart} at the leading order $\lambda$ read\\
\begin{subequations}\label{firstorder}
\begin{equation}
\Big[ r^2 \Big( \frac{d^2}{dr^2}+\frac{\gamma_1^2}{r^2} \Big) -n(n+1) \Big] l_+ u_n (r)=0
\end{equation}
\begin{equation}
\Big[ r^2 \Big( \frac{d^2}{dr^2}+\frac{\gamma_2^2}{r^2} \Big) -n(n+1) \Big] l_+ v_n (r)=0
\end{equation}
\end{subequations}
where $\gamma_1$ and $\gamma_2$ are the TM and TE (respectively) anisotropic factor given by\\
\begin{subequations}
\begin{eqnarray}
\gamma_1^2 &= & k_0^2\varepsilon_+\Big\{ 1-\lambda \Big[ 1-\frac{2m^2}{n(n+1)} \Big] \Big\} \nonumber \\ \gamma_2^2 & = &
k_0^2\varepsilon_+\Big\{ 1-\lambda \Big[ \frac{1-4m^2}{4n(n+1)-3}+\frac{2m^2}{n(n+1)} \Big] \Big\}\nonumber
\end{eqnarray}
\end{subequations}
Eqs. \eqref{firstorder} have the same structure of the radial equation for the isotropic case \cite{ref23}. The only difference is
the presence of the $\gamma_i$ terms that modify the arguments of the Riccati-Bessel functions and the quasi-potentials can be
written as\\
\begin{subequations}
\begin{equation}
U_{nm}(r,\theta,\varphi)=C_{int/ext}z_{\nu}(x)Y_{nm}(\theta,\varphi)
\end{equation}
\begin{equation}
V_{nm}(r,\theta,\varphi)=C_{int/ext}z_{\nu}(x)Y_{nm}(\theta,\varphi)
\end{equation}
\end{subequations}
where $z_{\nu}(x)$ corresponds to $j_{\nu}(x)$ inside the sphere and to $h_{\nu}^{(1)}(x)$ outside the sphere. Note also that
inside the sphere, where the anisotropy exists, $x=\gamma_i  r$, while outside the sphere $x=k_0 r$ (the surrounding medium is
still isotropic). Substituting these expressions in Eqs. \eqref{TMfield} and \eqref{TEfield} we obtain all the components of the
electric and magnetic fields in an uniaxial anisotropic spherical resonator\\
\begin{subequations}\label{quasiTMComponents}
\begin{eqnarray}
E_r &= & \frac{n(n+1)}{r^2} \Big[ \sqrt{\frac{\pi}{2}}z_{\nu}(\gamma_1 x)Y_{n,m}(\theta,\varphi) \Big],\\ rE_{\theta} &= &
-\frac{\partial^2}{\partial r\partial\theta} \Big[ \sqrt{\frac{\pi}{2}}z_{\nu}(\gamma_1 x)Y_{n,m}(\theta,\varphi) \Big]\\
rE_{\varphi} &= & \frac{1}{\sin\theta}\frac{\partial^2}{\partial r\partial\varphi} \Big[ \sqrt{\frac{\pi}{2}}z_{\nu}(\gamma_1
x)Y_{n,m}(\theta,\varphi) \Big]\\
 H_r &= & 0\\
rH_{\theta} &= & -\frac{ik_0\varepsilon_{\perp}}{\sin\theta}\frac{\partial}{\partial\varphi} \Big[
\sqrt{\frac{\pi}{2}}z_{\nu}(\gamma_1 x)Y_{n,m}(\theta,\varphi) \Big]\\ rH_{\varphi} &= &
ik_0\varepsilon_{\perp}\frac{\partial}{\partial\theta}  \Big[ \sqrt{\frac{\pi}{2}}z_{\nu}(\gamma_1 x)Y_{n,m}(\theta,\varphi)
\Big]\\
\end{eqnarray}
\end{subequations}
for quasi-TM fields. Similar expressions can be written for the quasi-TE fields by replacing $\gamma_1$ with $\gamma_2$,
exchanging the role of the electric and magnetic field and setting $\varepsilon_{\perp}=1$.

The characteristic equation can be found by applying the boundary conditions and it turns out to be
($\tilde{\gamma_i}=\gamma_i/k_0\sqrt{\varepsilon_+}$)\\
\begin{equation}\label{CEquasiTM}
\tilde{\gamma_1}\frac{[j_{\nu}(\gamma_1 kR)]'}{j_{\nu}(\gamma_1
kR)}=\frac{\varepsilon_{\perp}}{\sqrt{\varepsilon_+}}\frac{[h_{\nu}^{(1)}(k_0 R)]'}{h_{\nu}^{(1)}(k_0 R)}
\end{equation}
for the quasi-TM waves, and
\begin{equation}\label{CEquasiTE}
\tilde{\gamma_2}\frac{[j_{\nu}(\gamma_2 kR)]'}{j_{\nu}(\gamma_2 kR)}=\frac{1}{\sqrt{\varepsilon_+}}\frac{[h_{\nu}^{(1)}(k_0
R)]'}{h_{\nu}^{(1)}(k_0 R)}
\end{equation}
for the quasi-TE waves.\\ As can be seen from the previous equations, in the small anisotropy regime, the only effect of the
anisotropy is a rescaling of the radial coordinate; this is in accordance with the fact that an uniaxial crystal shows two
different refractive indexes: one in-plane ($\sqrt{\varepsilon_{xx}}$) and the other out-of-plane ($\sqrt{\varepsilon_{zz}}$).
Different refractive indexes correspond to different optical paths, and this is exactly reflected in the rescaling effect of the
anisotropy onto the radial part of the modes of the resonator. Note also that at this level of analysis, the anisotropy doesn't
affect the angular structure ($\theta$ and $\varphi$) of the modes. Another difference respect to the isotropic case is the value
of the coefficients on the right-hand side of the characteristic equations: while the coefficient for the quasi-TE wave is
analogous to its isotropic counterpart (if we substitute the isotropic dielectric constant $\varepsilon$ with the
anisotropy-averaged dielectric constant $\varepsilon_+$), the coefficient for the quasi-TM wave reveals the presence of the
anisotropy, since it is a ratio between the in-plane dielectric constant and the anisotropy-averaged one. This is not so
surprising because for a dielectric uniaxial crystal, only the TM component suffer directly anisotropy, while the TE component
does not, because the crystal is magnetical isotropic.
\section{IV. Whispering gallery Modes}
\subsection{A. Radial functions with large indices}
With the term whispering gallery mode (WGM) are commonly addressed the set of modes with a large index $n$; strictly speaking, the
real WGMs are only those for which it results that $n=m$ and the radial wavefunction shows no roots inside the resonator. However,
modes with indices $n\neq m$ and with $q > 1$, but close to unity, have properties that are close to those of WGMs: this means
that there is no great difference between a ``pure" WGM and other modes with nearest indices.

\begin{figure}[!t] 
 \includegraphics[width=0.5\textwidth]{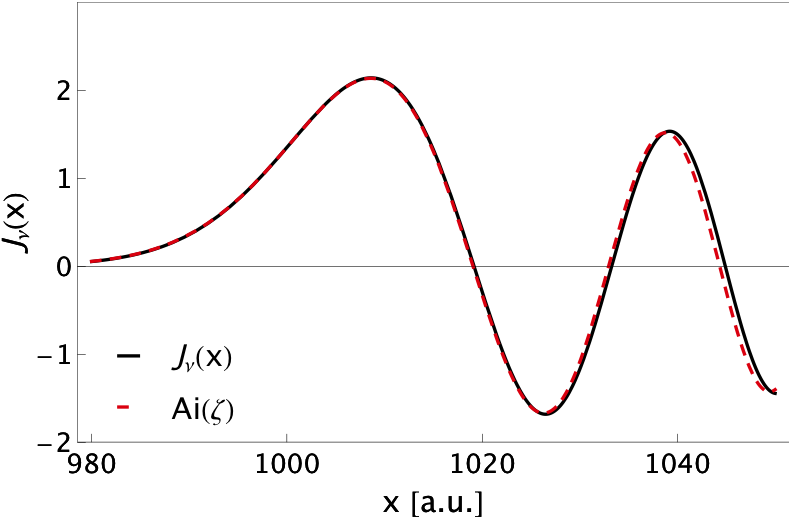} 
 \caption{(color online)
Comparison between $j_{\nu}(x)$ (solid black line) and its Airy approximation from Eq \eqref{Approx_a} (dashed red line) for
$\nu=1000.5$. The approximation holds very well up to $x\simeq\nu$, while for $x$ larger than $\nu$ (say for $x>1020$) it starts
to fail.} 
\label{figuraApprox1} 
\end{figure}

\begin{figure}[!t] 
\includegraphics[width=0.5\textwidth]{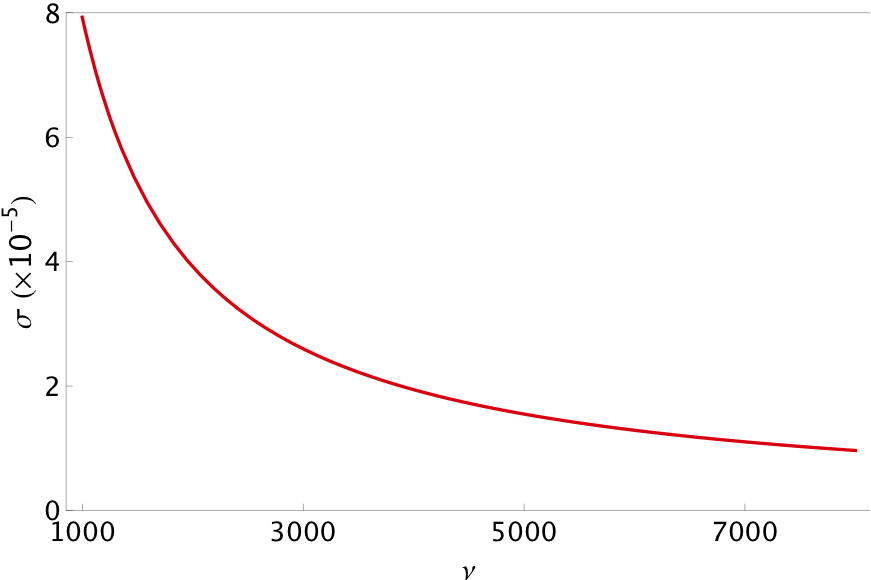}
\caption{(color online) The figure shows the accuracy $\sigma$ as a function of the Bessel index $\nu$; the accuracy is defined
as the ration between the difference of the true function $j_{\nu}(\nu)$ and its Airy approximation $\mathrm{Ai}(\zeta)$ and their
sum, i.e. $\sigma=[j_{\nu}(\nu)-\mathrm{Ai}(\zeta^*)]/[j_{\nu}(\nu)+\mathrm{Ai}(\zeta^*)]$, where $\zeta^*$ is $\zeta$ evaluated
for $x=\nu$. As can be seen, as $\nu$ grows, the approximation becomes more precise.} 
\label{compareApprox}
\end{figure}

\begin{figure}[!t] 
\includegraphics[width=0.5\textwidth]{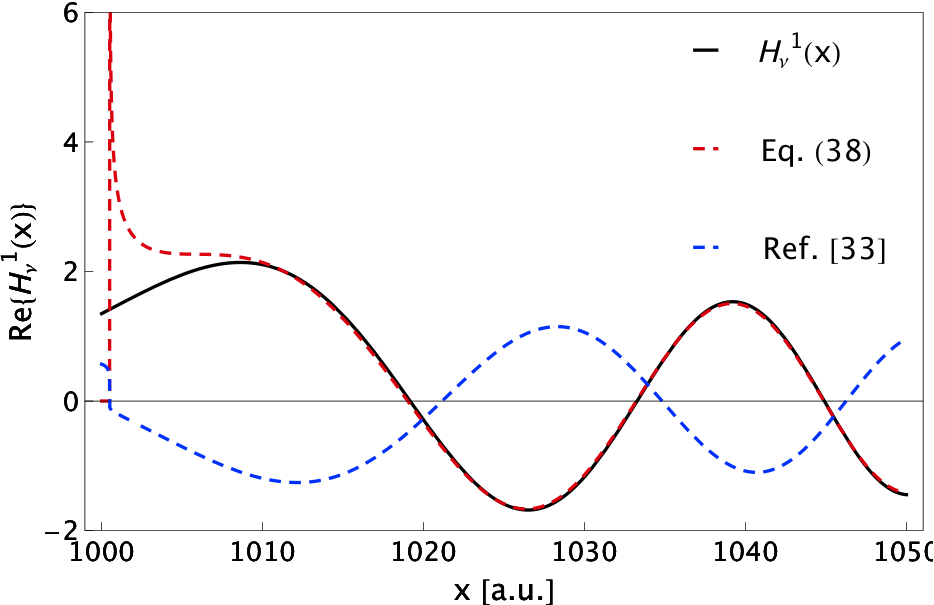} 
\caption{(color online) Comparison between the real part of $h_{\nu}^{(1)}(x)$ (black solid line), Eq \eqref{Happrox} (dashed red line) and Eq. (31) of
Ref.\cite{ref26} (dashed blue line) for $\nu=1000.5$. Our approximation works very well in the region in which the argument is
grater than the index (i.e. $x>1010$), while the approximation presented in Ref.\cite{ref26} is out of phase respect to the Hankel
function.} 
\label{approxH_real} 
\end{figure}

\begin{figure}[!t]
\includegraphics[width=0.5\textwidth]{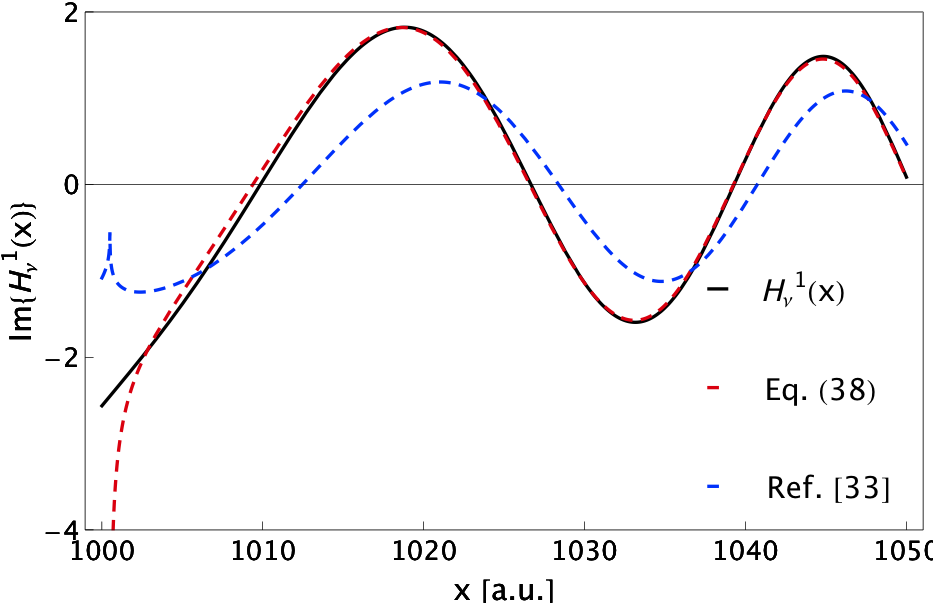} 
\caption{(color online) Same as Fig. \ref{approxH_real} but the
comparison is made for the imaginary part.} 
\label{approxH_imag} 
\end{figure}

\begin{figure}[!t]
\includegraphics[width=0.5\textwidth]{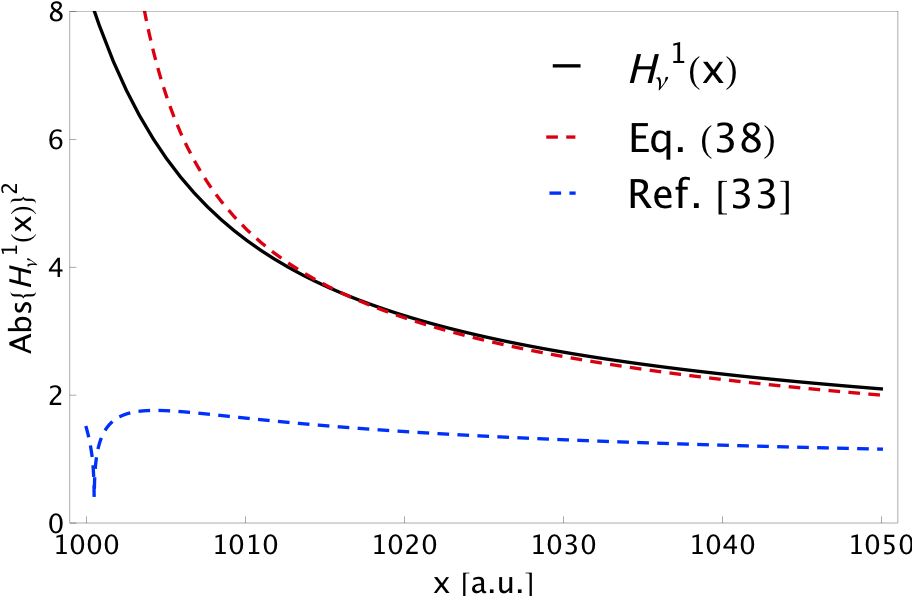} 
\caption{(color online) Same as Fig. \ref{approxH_real} but the
comparison is made for the absolute value; note in this case how the approximation presented in Ref.\cite{ref26} completely fails
to approximate the Hankel function.} 
\label{approxH_abs} 
\end{figure}

To study such modes, the first thing we have to do is to find a suitable approximation of Riccati-Bessel functions for large
index. This approximation is useful either from the numerical (where computing Bessel functions of large index is highly
time-consuming) or analytical (where the approximation gives the possibility to work with easier functions that suit better onto
the problem) point of view. The appropriate approximation, however, should be searched bearing in mind that the argument of the
Bessel function for a WGM near the sphere surface is of the order of its index, i.e. $\nu/x \simeq 1$. By introducing the
following change of variables\\
\begin{equation}
\zeta=\Big( \frac{2}{\nu} \Big) ^{1/3}(\nu-x)\nonumber,
\end{equation}
the Bessel function inside the dielectric resonator can be very well approximated by the Airy function of the first kind Ai as
follows \cite{ref31}:\\
\begin{subequations}\label{besselApprox}
\begin{eqnarray}
j_{\nu}(x) & \simeq & \sqrt{2} \Big( \frac{\nu}{2} \Big) ^{1/6} \textrm{Ai}(\zeta) ,\label{Approx_a}\\ \frac{d}{dx}[j_{\nu}(x)] &
\simeq & - \sqrt{2} \Big( \frac{2}{\nu} \Big) ^{1/6} \frac{d}{d\zeta} [\textrm{Ai}(\zeta)].
\end{eqnarray}
\end{subequations}
The accuracy of this approximation is of the order $\nu^{-1}$; if $\nu$ exceeds 1000, this accuracy is very satisfactory for many
calculations. this can be seen in Fig.\ref{figuraApprox1}, where Bessel functions of high order are compared with their Airy
approximation and in Fig.\ref{compareApprox}, where is shown that as $\nu$ grows, the accuracy of the approximation became
satisfactory.\\
For the solution outside the resonator (the Hankel function of the first kind) various approximations are available. Here we use the
following \cite{ref32}:\\
\begin{eqnarray}\label{Happrox}
h_{\nu}^{(1)}(x) & = & f(\eta)= \simeq  \frac{e^{i[\nu(\tan\eta-\eta)-\frac{\pi}{4}]}}{\sqrt{\frac{\pi}{2}\tan\eta}} \Big\{
1+\nonumber \\ & - &\frac{i}{\nu} \Big( \frac{1}{8\tan\eta}+\frac{5}{24}\frac{1}{\tan^3\eta} \Big)+ O[\nu^{-2}] \Big\}
\end{eqnarray}
where $\cos\eta=\nu/x$; if $\nu$ is large enough (heuristically $\nu>1000$) the imaginary term inside the curly brackets can be
neglected. The choice of this approximation rather than the one presented in Ref.\cite{ref26} reside in the fact that while the
former is very good when the argument of the Hankel function is greater than the index (that is precisely the case of the outer
functions), the latter is not suitable in this region, either for being out of phase with respect to Hankel function (as shown in
Figs. \ref{approxH_real} and \ref{approxH_imag}) or to not approximate in the correct way the original function (Fig.
\ref{approxH_abs}). Moreover, Fig. \ref{approxH_abs} shows that the field outside the resonator has all the characteristics of an
evanescent wave, i.e. it decays exponentially as the distance from the resonator surface grows.

In order to justify this evanescent behavior outside the resonator, one can directly solve Eq.\eqref{besselRadiale} in the limit
$r>R$ (but still close to the resonator surface), where the terms $x=k_0 r$ outside the derivation symbol can be substituted with
$k_0 R$, leading to the following equation:\\
\begin{equation}\label{evanescent}
\frac{d^2Z}{dx^2}+\frac{1}{k_0 R}\frac{dZ}{dx}+\Big[1-\frac{\nu^2}{(k_0 R)^2}\Big]Z=0,
\end{equation}
whose solution is:\\
\begin{equation}\label{evanescentSolution}
Z_{\nu}(x)=C_0e^{-\delta x},
\end{equation}
where\\
\begin{equation}
\delta=\Bigg[\sqrt{\Big(\frac{\nu}{k_0R}\Big)^2-1+\frac{1}{4(k_0R)^2}}-\frac{1}{2k_0 R}\Bigg].\nonumber
\end{equation}
This is, how we are expecting, the expression of an exponentially decreasing field, that is in perfect agreement with the
hypothesis that the field outside the resonator is evanescent due to total internal reflection.

It can be moreover noted that the oscillatory behavior of the field components outside the resonator (as depiscted in Figs.
\ref{approxH_real} and \ref{approxH_imag} for the radial component of the electric field) is not in contrast with this hypothesis,
since it only represent the behavior of the Hankel function as $r\rightarrow\infty$, i.e. it behaves like a runaway wave whose
intensity is decreasing as $1/r^2$. In the case of WGM, however, no radiation will run away towards infinity since the external
field is evanescent, i.e. the radiation is trapped inside the WGM and rapidly decreases toward zero when the field goes outside
the resonator. 
\subsection{B. Angular functions with large indices}
For large indices $n$, the WGM field is concentrated in a narrow interval of angles $\theta$ near $\theta_0=\pi/2$; this makes
possible to approximate the associated Legendre functions (i.e. the $\theta$-part of the scalar spherical harmonics) with large
indices , with Hermite polynomials with small indices as follows:\\
\begin{equation}\label{Yapprox}
Y_{nm}(\theta,\varphi) \simeq \frac{\sqrt{m}}{2^{w}\sqrt{\pi}w !}H_{w}(\sqrt{m}\alpha)e^{-\frac{m}{2}\alpha^2}e^{im\varphi}.
\end{equation}
Detailed calculations for obtaining this result are shown in Appendix B.\\ %
\subsection{C. Roots of characteristic equations}
The approximations exploited in the previous section are very useful in finding an analytical solution to the characteristic
equation for the eigenfrequencies of the resonator; however, due to the anisotropy, some changes in the definition of the
variables used above must be done. First of all, the $x$ appearing in Eqs. \eqref{besselApprox} and \eqref{Happrox} has to be
different for the inner and outer functions, due to the fact that the anisotropy is confined only inside the resonator; we can
then define $x=k_0R$ as the outer variable and by consequence the inner variable results to be
$y=\tilde{\gamma_i}\sqrt{\varepsilon_+}x$. Then, the definition of $\zeta$ must be changed into $\zeta=(2/\nu)^{1/3}(\nu-y)$.
After that, by substituting Eqs. \eqref{besselApprox} and \eqref{Happrox} into Eq \eqref{CEquasiTM}, the characteristic equation
for quasi-TM field gives \cite{ref33}\\
\begin{eqnarray}\label{eigenWGM}
\tilde{\gamma_1} \frac{1}{\textrm{Ai}(\zeta)}\frac{d\textrm{Ai}(\zeta)}{d\zeta}= \frac{\varepsilon_{\perp}}{\sqrt{\varepsilon_+}}
\Big( \frac{\nu}{2} \Big) ^{1/3}\times \nonumber \\ \times \Bigg[ \frac{1}{4} \Big( \frac{2x}{x^2-\nu^2} \Big) -
i\sqrt{1-\frac{\nu^2}{x^2}} \Bigg],
\end{eqnarray}

the equation for the quasi-TE field can be deduced by this one upon changing $\tilde{\gamma_1}$ with $\tilde{\gamma_2}$ and
putting $\varepsilon_{\perp}=1$. \\ In order to find an approximate formula for the solutions of this equation, let us firstly
analyze the limiting case in which $\nu\rightarrow\infty$; in this case the right-hand side of the equation goes to infinity and
the only possible solution is that $\textrm{Ai}(\zeta)=0$, whose solutions are the zeros of the Airy function $\zeta_q$. Let us
denote with $\Delta\zeta_q$ the first order correction to these roots; expanding both left-hand and right-hand size of Eq
\eqref{eigenWGM} in power series with a first order accuracy to terms $\Delta\zeta_q$ we can obtain the first order correction to
the roots $\zeta_q$, whose expression is\\
\begin{equation}\label{correction}
\Delta\zeta_q=\frac{\tilde{\gamma_1}\sqrt{\varepsilon_+}}{\varepsilon_{\perp}\alpha} \Big( \frac{2}{\nu} \Big) ^{1/3},
\end{equation}
where\\
\begin{equation}
\alpha=\frac{x_q}{2(x_q^2-\nu^2)}-i\sqrt{1-\frac{\nu^2}{x_q^2}},
\end{equation}
and $x_q$ is obtained by substituting the value of the first zero of the Airy function ($\zeta_q=-2.33811$) into the definition of
$\zeta$ and inverting that relation with respect to $x$.\\ Taking into account the definition of $\zeta$, the eigenvalues of the
wave numbers for the anisotropic resonator can be represented in the following explicit form\\
\begin{equation}
k_{0q}=\frac{\nu- \Big( \frac{2}{\nu} \Big) ^{1/3} (\zeta_q+\Delta\zeta_q)}{\tilde{\gamma_1}\sqrt{\varepsilon_+}R}.
\end{equation}
Note that because the quantity $\Delta\zeta_q$ is complex, the eigenvalues of the wave number are also complex. The real part of
the wave number then determines the eigenfrequencies of the mode. Complex eigenfrequencies are fully compatible with the open
cavity. As can be seen from Eq. \eqref{correction}, this approximation has an accuracy of $\nu^{-1/3}$. more accurate asymptotic
expressions that allow the calculation of the positions of resonances of the modes in an isotropic dielectric spherical resonator
have been largely studied in literature (see for example Ref.\cite{ref34,ref35,ref36,ref37,ref38,ref39} and references therein)
and they were given with various accuracy with respect to the index $\nu$; in Ref.\cite{ref39} analytic calculations are carried
out to the order $\nu^{-1/3}$, in Ref.\cite{ref35} the eigenfrequencies are calculated with an accuracy of $O[\nu^{-2/3}]$, while
in Ref.\cite{ref38} the authors give an expression up to the order $O[\nu^{-8/3}]$. Here we report the anisotropic correction of
the formula found in Ref.\cite{ref34} that gives the eigenfrequencies with a precision of the order of $O[\nu^{-1}]$\\
\begin{eqnarray}
\tilde{\gamma_i} x_{\nu}^{(q)} & = & \Big\{ \nu - \Big( \frac{\nu}{2} \Big) ^{1/3}\zeta_q -
\sqrt{\frac{\varepsilon_+}{\varepsilon_+ -1}}P +\nonumber \\
 & + &\frac{3}{10} \Big( \frac{1}{4\nu} \Big) ^{1/3} \zeta_q^2 - \Big( \frac{1}{2\nu^2} \Big) ^{1/3} \Big(
 \frac{\varepsilon_+}{\varepsilon_+ -1} \Big) ^{3/2}\times\\
& \times & P \Big(\frac{2}{3}P^2-1 \Big) \zeta_q\nonumber  +O[\nu^{-1}] \Big\},
\end{eqnarray}
where $P=1/(\tilde{\gamma_1}\varepsilon_+)$ for quasi-TM modes and $P=\varepsilon_{\perp}/ (\tilde{\gamma_2}\varepsilon_+)$ for
quasi-TE modes.\\ 
\subsection{D. Whispering Gallery Modes}
We now have all the elements for writing the explicit expressions for the radial, polar and azimuthal components of the quasi-TM
and quasi-TE WGMs. Taking the approximations \eqref{besselApprox}, \eqref{Happrox} and \eqref{Yapprox}, the equations for the
components of the quasi-TM fields defined in Eqs. \eqref{quasiTMComponents} become\\
\begin{subequations}\label{quasi_TM_WGM_in}
\begin{eqnarray}
E_r & = &\frac{n(n+1)\sqrt{m}}{r^2 2^{w}\sqrt{\pi}w ! }\Big( \frac{\nu}{2} \Big) ^{1/6} \textrm{Ai}(\zeta)\times\nonumber \\ &
\times & \mathrm{H}_{w}(\sqrt{m}\alpha)e^{-\frac{m}{2}\alpha^2}e^{im\varphi},\\ rE_{\theta} & = & \tilde{\gamma_1}
k_0\frac{\sqrt{2\varepsilon_+ m}}{2^{w}w !}\Big( \frac{2}{\nu} \Big) ^{1/6}\frac{d\textrm{Ai}(\zeta)}{d\zeta}\times \nonumber \\ &
\times & \frac{\partial}{\partial\theta} \Big[\mathrm{H}_{w}(\sqrt{m}\alpha)e^{-\frac{m}{2}\alpha^2} \Big] e^{im\varphi},\\
rE_{\varphi} & = & \frac{im^{3/2}\tilde{\gamma_1} k_0}{\sin\theta2^{w-1/2}w !}\Big( \frac{2}{\nu} \Big) ^{1/6}
\frac{d\textrm{Ai}(\zeta)}{d\zeta}\times \nonumber \\ & \times &
\mathrm{H}_{w}(\sqrt{m}\alpha)e^{-\frac{m}{2}\alpha^2}e^{im\varphi},
\end{eqnarray}
\begin{eqnarray}
H_r & = & 0,\\ rH_{\theta} & = & \frac{k_0\varepsilon_{\perp}m^{3/2}}{\sin\theta 2^{w}w !} \Big( \frac{\nu}{2} \Big)
^{1/6}\textrm{Ai}(\zeta)\times \nonumber \\ & \times & \mathrm{H}_{w}(\sqrt{m}\alpha)e^{-\frac{m}{2}\alpha^2}e^{im\varphi},\\
rH_{\varphi} & = & \frac{ik_0\varepsilon_{\perp}\sqrt{m}}{2^{w}w !} \Big( \frac{\nu}{2} \Big) ^{1/6}\textrm{Ai}(\zeta)\times
\nonumber \\ & \times & \frac{\partial}{\partial\theta} \Big[ \mathrm{H}_{w}(\sqrt{m}\alpha)e^{-\frac{m}{2}\alpha^2} \Big]
e^{im\varphi}.
\end{eqnarray}
\end{subequations}
for the field inside the resonator, while for the field outside the resonator the expressions are the following:\\
\begin{subequations}\label{quasi_TM_WGM_out}
\begin{eqnarray}
E_r & = &\frac{n(n+1)\sqrt{m}}{r^2 2^{w}\sqrt{\pi}w ! }f(\eta)\times\nonumber \\ & \times &
\mathrm{H}_{w}(\sqrt{m}\alpha)e^{-\frac{m}{2}\alpha^2}e^{im\varphi},\\ rE_{\theta} & = & \tilde{\gamma_1}
k_0\frac{\sqrt{2\varepsilon_+ m}}{2^{w}w !}\Big(\frac{1}{\nu\sin\eta}\Big)\frac{df(\eta)}{d\eta}\times \nonumber \\ & \times &
\frac{\partial}{\partial\theta} \Big[\mathrm{H}_{w}(\sqrt{m}\alpha)e^{-\frac{m}{2}\alpha^2} \Big] e^{im\varphi},\\ rE_{\varphi} &
= & \frac{im^{3/2}\tilde{\gamma_1} k_0}{\sin\theta2^{w-1/2}w !}\Big(\frac{1}{\nu\sin\eta}\Big)\frac{df(\eta)}{d\eta}\times
\nonumber \\ & \times & \mathrm{H}_{w}(\sqrt{m}\alpha)e^{-\frac{m}{2}\alpha^2}e^{im\varphi},
\end{eqnarray}
\begin{eqnarray}
H_r & = & 0,\\ rH_{\theta} & = & \frac{k_0\varepsilon_{\perp}m^{3/2}}{\sin\theta 2^{w}w !}f(\eta)\times \nonumber \\ & \times &
\mathrm{H}_{w}(\sqrt{m}\alpha)e^{-\frac{m}{2}\alpha^2}e^{im\varphi},\\ rH_{\varphi} & = &
\frac{ik_0\varepsilon_{\perp}\sqrt{m}}{2^{w}w !}f(\eta)\times \nonumber \\ & \times & \frac{\partial}{\partial\theta} \Big[
\mathrm{H}_{w}(\sqrt{m}\alpha)e^{-\frac{m}{2}\alpha^2} \Big] e^{im\varphi}.
\end{eqnarray}
\end{subequations}

Similar expressions can be found for quasi-TE WGMs by interchanging the roles of the electric and magnetic field in the previous
expressions, changing $\tilde{\gamma_1}$ with $\tilde{\gamma_2}$, putting $\varepsilon_{\perp}=1$ and changing $i$ with $-i$.\\

\begin{figure}[!t] 
\includegraphics[width=0.5\textwidth]{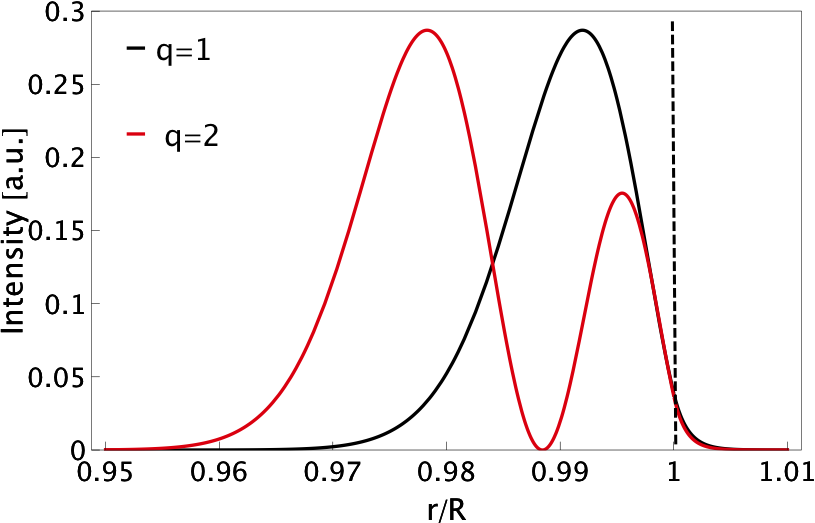} 
\caption{(color online) Radial part of the fundamental (black line) and first excited (red line) WGM for the anisotropic resonator; the vertical
dashed line indicates the position of the resonator surface.  The fundamental mode has indices $n=m=1000$ and $q=1$, while the
first excited mode has the same $n$ and $m$ indices but $q=2$.} 
\label{WGM_fundamental_radial} 
\end{figure}

\begin{figure}[!t] 
\includegraphics[width=0.5\textwidth]{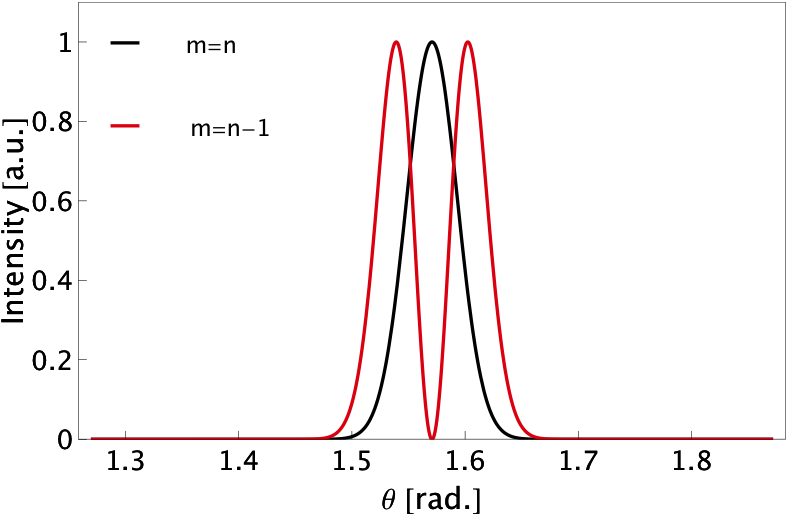} 
\caption{(color online) Polar part of the fundamental (black line) and first polar-excited  (red line) WGM for the anisotropic resonator. The fundamental mode has indices $n=m=1000$ and $q=1$, while the first polar-excited mode has $n=1000$, $m=n-1$ and $q=1$.} 
\label{WGM_fundamental_polar} 
\end{figure}

Figures \ref{WGM_fundamental_radial} and \ref{WGM_fundamental_polar} show the behavior of the fundamental quasi-TM radial (no
nodes in radial direction, i.e. $q=1$) and polar ( i.e. $n=m$) WGM component $r^2E_r$ and the ``first excited" radial ($q=2$) and
polar ($m=n+1$) mode for the same component of the quasi-TM field; the physical parameters have been set to be
$\varepsilon_{xx}=5.30$, $\varepsilon_{zz}=6.47$ ($\mathrm{LiNbO_3}$) and $\lambda=1064$ nm. Note that the radial component has
its maximum very close to the sphere surface (dashed vertical line in Fig. \ref{WGM_fundamental_radial}), and its position shifts
on the left, i.e. on the inner part of the resonator as the radial number $q$ increases. The polar part, instead, is localized
around $\theta=\pi/2$ in its fundamental state and, as $m$ becomes smaller than $n$, the maxima of the polar component tent to
repel each other from $\theta=\pi/2$. 

\begin{figure}[t] 
\includegraphics[width=0.5\textwidth]{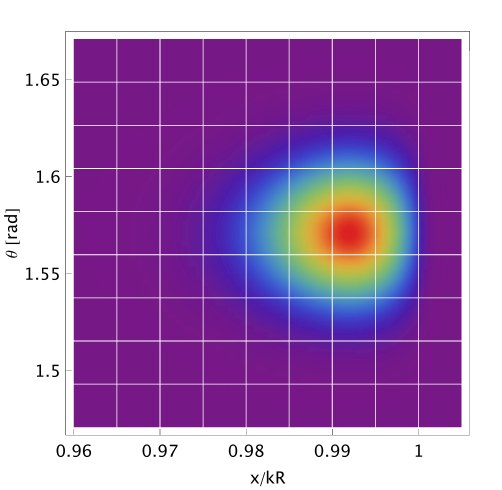} 
\caption{(color online) Intensity distribution of the electric
field of the fundamental quasi-TM WGM. The WGM quantum numbers are $n=m=1000$,$q=1$.} 
\label{TF1} 
\end{figure}

\begin{figure}[t] 
\includegraphics[width=0.5\textwidth]{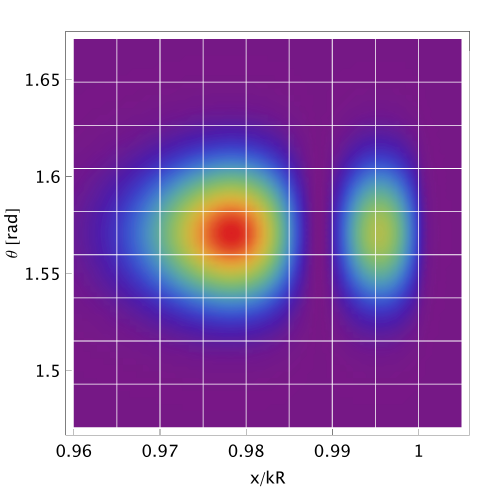} 
\caption{(color online) Same as Fig. \ref{TF1} but for $q=2$; this mode represents the first radially excited WGM.} 
\label{TF2} 
\end{figure}

\begin{figure}[t] 
\includegraphics[width=0.5\textwidth]{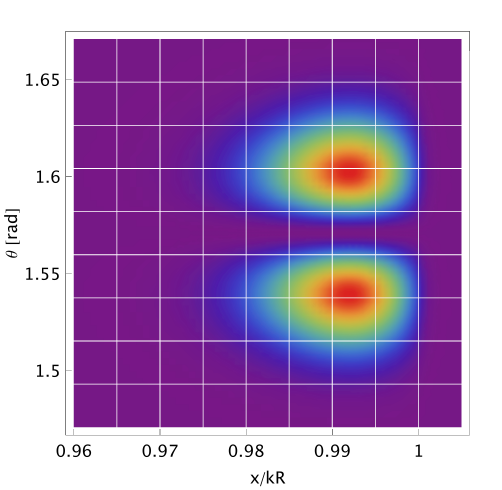} 
\caption{(color online) Same as Fig. \ref{TF1} but for $n-m=1$;
this mode represents the first polar excited WGM.} 
\label{TF3} 
\end{figure}

\begin{figure}[!h] 
\includegraphics[width=0.5\textwidth]{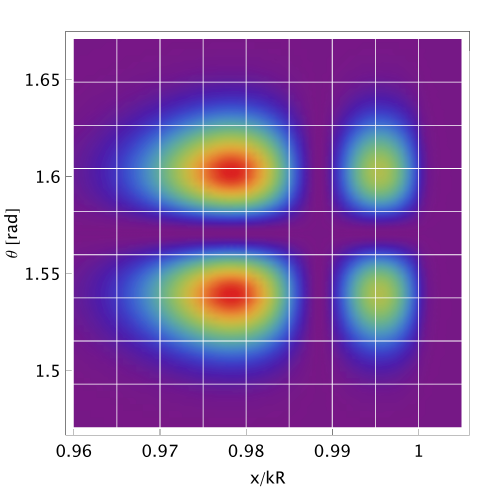} 
\caption{(color online) Same as Fig. \ref{TF1} but for $n-m=1$ and $q=2$; this mode represents the first radially and polar excited WGM.}
\label{TF4} 
\end{figure}

In figures \ref{TF1} to \ref{TF4} the intensity distribution of the \emph{total} electric field of a quasi-TM (i.e.
$\mathbf{E}_{TM}=E_r\hat{r}+E_{\theta}\hat{\theta}+E_{\varphi}\hat{\varphi}$) is shown, where the components $E_k$ ($k=r, \theta,
\varphi$) are given by Eqs. \eqref{quasi_TM_WGM_in} for the field inside the resonator and Eqs. \eqref{quasi_TM_WGM_out} for the
field outside the resonator; $\hat{r}$,$\hat{\theta}$ and $\hat{\varphi}$ represent the unit vectors of the spherical basis
($r$,$\theta$,$\varphi$). In order to obtain the intensity distribution of such a field, one has to sum the square modulus of each
component of the electric field; however, in this particular case, the contribution of $E_{\theta}$ and $E_{\varphi}$ is very
small and localized at the resonator surface, and the total field is, with a good level of approximation, fully determined by its
radial component. The intensity distribution for the magnetic field components of a quasi-TM mode can be straightforwardly
obtained from Eqs. \eqref{quasi_TM_WGM_in} and \eqref{quasi_TM_WGM_out} or by nothing that the $H_{\theta}$ component of the
magnetic field has the same intensity distribution as the radial electric field component $E_r$ and the $H_{\varphi}$ component,
because of the presence of the derivative with respect to $\theta$, has the same intensity distribution as the one depicted in
Fig. \ref{TF3}.

As the reader can see from these figures, the field is nonzero even after the resonator surface ($x=1$); this is not surprising
because in this region the total field is evanescent due to the fact that it has been total internal reflected by the resonator,
i.e. the field is confined in the resonator WGM.

\section{Conclusions}
In this work, we have developed a classical-optics theory for an uniaxial spherical whispering gallery resonator. We have presented and discussed the mode structure in the limit of small anisotropy for such resonator,
and obtained its spectrum. Moreover, we have furnished a thorough discussion on the boundary conditions and asymptotic expressions for the electromagnetic field in WGRs. Our results may be easily generalized to achieve a quantum theory of WGRs.

\section{Acknowledgements}
The authors want to thank Josef F\"{u}rst, Christoph Marquardt and Dmitry Strekalov for fruitful discussions.

\section{Appendix A: coefficients of Eqs. (22)}
Here are reported the explicit expressions of the coefficients that appear on Eqs. \eqref{radialpart}. In order to express them in
a compact form, let us introduce the following quantities:\\
\begin{subequations}
\begin{eqnarray}
g_{\pm} & = & k_o^2\varepsilon_{\pm},\nonumber \\ f_n & = & \frac{1}{2n+2},\nonumber \\ T_n & = & r^2 \hat{l}_x -n(n+1),\nonumber
\\ l_+ & = & \frac{\partial^2}{\partial r^2}+k_0^2\varepsilon_+.\nonumber
\end{eqnarray}
\end{subequations}
With these parameters defined, the $a$ and $b$s coefficients of Eq. \eqref{radialpart} become:\\
\begin{subequations}
\begin{eqnarray}
a_{1,n}^{TM}  & = & T_n \Big[l_+ - \frac{1-4m^2}{4n(n+1)-3}g_- \Big]+2g_- m^2,\nonumber\\ a_{2,n}^{TM} & = & 2g_-
(n-m+1)(n-m+2)f_n f_{n+1}T_n,\nonumber\\ a_{3,n}^{TM} & = & 2g_- (n+m)(n+m-1)f_n f_{n-1} T_n,\nonumber\\ b_{1,n} & = & f_n (n-m+1)
\Big[ r \hat{l}_x -(n+2)\frac{d}{dr} \Big], \nonumber\\ b_{2,n} & = & f_n (n+m) \Big[ r \hat{l}_x + (n-1)\frac{d}{dr} \Big],
\nonumber
\end{eqnarray}
\begin{eqnarray}
a_{1,n}^{TE} & = & \Big\{ \Big[ r^2\frac{d^2}{dr^2}-n(n+1) \Big] \Big[ 1+\lambda \Big(\frac{1-4m^2}{4n(n+1)-3} \Big) \Big]
+\nonumber \\
 & + & (1-\lambda)g_+ r^2 \Big\} \hat{l}_x - 2m^2g_- (1-\lambda),\nonumber\\
a_{n,2}^{TE} & = & 2f_n f_{n+1} (n-m+1)(n-m+2)\times\nonumber \\ & \times & \Big[ r^2\frac{d^2}{dr^2}-(2n+3)r\frac{d}{dr}+n(n+1)
\Big],\nonumber\\ a_{n,3}^{TE} & = & 2f_n f_{n-1} (n+m)(n+m-1)\times\nonumber \\ & \times & \Big[
r^2\frac{d^2}{dr^2}-(2n-1)r\frac{d}{dr}+(n+1)(n-3) \Big].\nonumber
\end{eqnarray}
\end{subequations}
\section{Appendix B: approximation of scalar spherical harmonics for large indices}
The equation for the $\theta$-part of spherical harmonics is the following\\
\begin{equation}
\frac{1}{\sin\theta}\frac{d}{d\theta} \Big( \sin\theta\frac{df}{d\theta} \Big) + \Big[ n(n+1) - \frac{m^2}{\sin^2\theta}
\Big]f=0,\nonumber
\end{equation}
whose solutions are the associated Legendre functions $f(\theta)=P_n^m(\cos\theta)$. Since WGMs are located near the equator of
the resonator, the correspondent functions $f(\theta)$ will be peaked near the angle $\theta_0=\pi/2$; in order to find an
approximate expression for the polar part of the spherical harmonics, let us introduce the new variable $\alpha=\pi/2-\theta$:
substituting into equation above gives\\
\begin{equation}
\frac{d^2}{d\alpha^2} - \tan\alpha\frac{df}{d\alpha} + \Big[ n(n+1) - \frac{m^2}{\cos^2\alpha} \Big] f=0.\nonumber
\end{equation}

We note that, since the modes are localized near the equator, $\alpha\ll 1$ and this allow us to expand in power series the trigonometric functions that appear
in the previous equation, i.e. $\tan\alpha\simeq\alpha$ and $1/\cos^2\alpha\simeq 1+\alpha^2$. Substituting in the previous
equation, writing $f(\alpha)=G(\alpha)e^{\alpha^2/4}$ and performing the change of variables $\alpha=x/[(m^2+1/4)^{1/4}]=\xi x$ we
obtain\\
\begin{equation}
\frac{d^2G}{dx^2} + \Big\{ \xi^2 \Big[ n(n+1) - m^2 \Big] -x^2 \Big\} G=0,\nonumber
\end{equation}
Introducing the quantity $w=n-m$ and remembering that WGMs are characterized by high values of the indices, i.e. $n,m\gg1$, the
first term that appears inside the curly brackets can be simplified as $2w+1$. With this substitution the last equation is
precisely the Hermite-Gauss equation, whose solutions have the form $G(x)\simeq H_{w}(x)e^{-x^2/2}$.  Function $f(\alpha)$ then
becomes\\
\begin{equation}
f(\alpha)= P_n^m(\cos\theta)\simeq NH_{w}(\sqrt{m}\alpha)e^{-\frac{m}{2}\alpha^2},\nonumber
\end{equation}
where $N$ is a normalization factor whose expression could be found by requiring that the norm of $f(\alpha)$ integrated over the
real axis is one. This equation gives the approximated form of the associated Legendre functions for WGMs; substituting it into
the definition of the scalar spherical harmonics gives exactly Eq \eqref{Yapprox}.



\begin{thebibliography}{99}
\bibitem{ref24} In this paper we consider a nonmagnetic resonator, i.e. $\mu=1$ surrounded by air ($\varepsilon_{ext}=1$).
\bibitem{ref28} In obtaining the azimuthal component of the quasi-TM magnetic field following relation was used: $\Big[
    \varepsilon_{\theta\theta}\frac{\partial}{\partial\theta} - \varepsilon_{r\theta}\frac{\partial}{\partial r} \Big( r \Big)
    \Big]\hat{l}_{\theta} = \hat{l}_{\theta} \Big[ \varepsilon_{\theta\theta}\frac{\partial}{\partial\theta} +
    \varepsilon_{r\theta} \Big(1-r\frac{\partial}{\partial r} \Big) \Big]$
\bibitem{ref30} Eq. (5) of Ref.\cite{ref29} contains an error: the expresison for $\cos(2\theta)Y_{n,m}(\theta,\varphi)$ is
    incorrect. Here we report the correct relation, in which the argument of the spherical harmonics is omitted for the sake of
    simplicity: $\cos(2\theta)Y_{n,m}=A_{n,m}Y_{n+2,m}+B_{n,m}Y_{n-2,m}+C_{n,m}Y_{n,m}$ with $A_{n,m}=2(n-m+1)(n-m+2)f_nf_{n+1}$,
    $B_{n,m}=2(n+m)(n+m+1)f_n f_{f-1}$ and $C_{n,m}=(1-4m^2)/[4n(n+1)-3]$
\bibitem{ref33} To write this equation we have used the expression in Eq \eqref{Happrox} stopped at the order $O[\nu^{-1}]$
\bibitem{ref40} As can be seen from Eqs. \eqref{TMcomponent}, the radial component of the TM field has a dependence on the radial
    wavefunction itself, via the second derivative of the potential. The tangential component, instead, depends on the derivative
    (with respect to the radial variable) of the wavefunction.
\end{thebibliography}
\end{document}